\documentclass[twocolumn]{aastex701}

\usepackage{enumitem}
\usepackage{amsmath}
\usepackage{comment}
\begin{document}

\title{Clump Migration in Disk Galaxies: Revisiting Chandrasekhar's Dynamical Friction}

\author[0009-0009-7497-3431]{Pushpak Pandey}
\affiliation{Inter-University Centre for Astronomy and Astrophysics, Ganeshkhind, Post Bag 4, Pune 411007, India}
\email{pushpak@iucaa.in}

\author[0000-0002-8768-9298]{Kanak Saha}
\affiliation{Inter-University Centre for Astronomy and Astrophysics, Ganeshkhind, Post Bag 4, Pune 411007, India}
\email{kanak@iucaa.in}

\begin{abstract}

Massive stellar clumps are thought to migrate toward galactic centers through dynamical friction, contributing to bulge growth and the structural evolution of disk galaxies. While the classical Chandrasekhar formalism is widely used to estimate clump inspiral times, its validity for extended, evolving clumps embedded in realistic galactic disks remains uncertain. We test this formalism using an isolated disk galaxy simulation by identifying and tracking individual stellar clumps over multiple orbital periods. We develop a position- and mass-based tracking algorithm that follows long-lived clumps and compare their measured orbital evolution with predictions from a dynamical friction model constructed from the simulated baryonic and dark matter mass distributions. After excluding the initial transient phase, we identify nine long-lived clumps, eight of which migrate inward while losing 60-90\% of their initial mass. The analytical model reproduces the overall dependence of migration on clump mass and galactocentric radius, but for six clumps it overestimates the inspiral time by factors of $\sim2$-$10$. Since mass loss would reduce the dynamical friction force, it cannot account for the observed faster migration. Our results suggest that additional physical processes, for example clump-clump interactions, non-circular orbits, and the time-dependent disk potential play an important role in regulating clump migration beyond the assumptions of the classical Chandrasekhar formulation.
\end{abstract}

\section{Introduction}
\label{sec:intro}
The morphological diversity of galaxies provides a fundamental probe of galaxy formation and evolution. Galaxies in the local Universe exhibit a wide range of structural properties, from ordered spiral and elliptical systems to irregular morphologies, with the galaxy population evolving significantly across cosmic time. A prominent feature of this evolution is the growing presence of massive star-forming clumps in high-redshift disk galaxies. High-resolution observations with the Hubble Space Telescope have revealed that galaxies at redshifts $z\sim0.5$--3 frequently exhibit irregular, clumpy morphologies dominated by kiloparsec-scale star-forming structures with stellar masses ranging from $10^{7}$ to $10^{9},M_{\odot}$ \citep{2005_Elmegreen,2007_Bournard_federic,2009Agertz_Teyssier,2011Genzel,2015Guo}. Similar, although generally less massive, clumpy structures are also observed in nearby spiral galaxies \citep{2006Elmegreen,2017Fisher}, suggesting that gravitational fragmentation of galactic disks is a recurring process whenever disks become sufficiently unstable due to high gas fractions, low velocity dispersion, or enhanced surface densities.

The formation of giant clumps is generally understood as a consequence of gravitational instabilities operating in turbulent, self-gravitating galactic disks \citep{1964toomre,1999Noguchi,2004Immeli_a,2004Immeli_b,2007_Bournard_federic,2009Dekel,Dekeletal2022}. Once formed, these massive clumps become important dynamical constituents of their host galaxies, redistributing mass and angular momentum through their interactions with the surrounding disk and dark matter halo. Their subsequent orbital evolution is therefore of considerable interest, as the inward migration of clumps has long been proposed as a key mechanism for driving secular galaxy evolution. In particular, the gradual inspiral of massive clumps can contribute to the growth of central bulges, enhance nuclear star formation through the accumulation of gas and stars in the inner regions, and potentially facilitate the fueling of central supermassive black holes \citep{1975Tremaine_Ostriker,1983Lin_Tremaine,1999Noguchi,2007_Bournard_federic,2009Dekel,2010caverino,2022Borgohain,2025Kataria,ZuYongdaetal2026}.

An important mechanism contributing to the orbital decay of massive clumps is gravitational dynamical friction. As a clump moves through the surrounding stellar and dark matter background, it gravitationally perturbs nearby particles, generating a density wake that trails its orbit \citep{TremaineWeinberg1984,Weinberg1986,Tamfaletal2021}. The gravitational attraction exerted by this wake acts as a drag force opposite to the direction of motion, leading to a gradual loss of orbital energy and angular momentum \citep{1943Chandrasekhar,2008Binney_tremaine}. Under the idealized assumptions of an infinite, homogeneous, and isotropic background with a Maxwellian velocity distribution, Chandrasekhar \citep{1943Chandrasekhar} derived an analytic expression for this drag force. Owing to its simplicity and predictive power, Chandrasekhar's formalism has become one of the fundamental tools for estimating orbital decay in a wide variety of astrophysical systems, including satellite galaxies, globular clusters, massive black holes, and stellar systems \citep{2008Binney_tremaine}.

Despite its widespread application, the assumptions underlying Chandrasekhar's derivation are only approximately satisfied in galactic disks. Unlike the idealized background assumed in the classical treatment, disk galaxies exhibit strong density gradients, ordered rotation, anisotropic velocity distributions, finite thickness, and time-dependent non-axisymmetric structures such as spiral arms and bars. Moreover, the clumps themselves are not rigid point-like perturbers. They can be extended, elongated, and irregular in morphology, while simultaneously evolving through tidal interactions, mass exchange, mergers, and, in gas-rich systems, stellar feedback \citep{1999Colpi_mayer,2003Mouri_Taniguchi,Dekeletal2023}. These departures from the assumptions of the classical theory complicate the generation of gravitational wakes and the associated transfer of orbital energy and angular momentum, raising important questions regarding the applicability of Chandrasekhar's formalism to clump migration in realistic galactic environments. Although numerous numerical studies have demonstrated that massive clumps migrate toward galactic centres on timescales of a few hundred Myr to a few Gyr \citep{2007_Bournard_federic,2010caverino,2015MTamburello,2017Mandelkar}, comparatively few have quantitatively evaluated how well the orbital evolution predicted by Chandrasekhar's dynamical friction prescription reproduces the migration measured in self-consistent galaxy simulations. Establishing the regime over which the classical formalism remains applicable, and identifying the conditions under which it breaks down, is therefore essential for interpreting clump migration in evolving galactic disks.

Theoretical estimates of clump migration are often based on simplified prescriptions derived from dynamical friction theory \citep{1943Chandrasekhar}, assuming idealized backgrounds such as isothermal halos, flat rotation curves, or a homogeneous and isotropic density and velocity distributions. While these approaches provide useful scaling relations for the dependence of migration timescales on clump mass and orbital properties, they do not capture the complex and evolving environments of realistic disk galaxies. In particular, massive clumps embedded within unstable, rotating disks interact with time-dependent non-axisymmetric structures, exchange angular momentum with both the stellar disk and dark matter halo, and experience a background potential that evolves as the galaxy undergoes structural transformation \citep{2010caverino,2008Binney_tremaine}. Furthermore, the applicability of analytic dynamical friction estimates is complicated by the intrinsic properties of the clumps themselves. Unlike idealized point-like perturbers assumed in classical treatments, giant clumps in simulations and observations often exhibit extended, elongated, and irregular morphologies \citep{Mandelkeretal2014}. Their evolving shapes, internal structures, and tidal distortions can modify the gravitational wake they generate and consequently influence the efficiency of angular momentum exchange with the surrounding medium \citep{Fujiietal2006}. Consequently, the applicability of Chandrasekhar's formalism to predicting the orbital decay of clumps in realistic disk environments remains an open question.

This motivates the use of high-resolution, self-consistent $N$-body simulations to quantify the efficiency of clump migration and assess the validity of analytic dynamical friction estimates in evolving disk environments. By focusing on a collisionless stellar system, such simulations provide a controlled framework for isolating the role of gravitational interactions and angular momentum exchange, while excluding additional physical processes associated with gas dynamics, star formation, and feedback. In this work, we compare analytically predicted and numerically measured clump migration in isolated disk galaxy simulations. We identify and track stellar clumps across multiple simulation snapshots and estimate the dynamical friction force using Chandrasekhar's formalism with the density, circular velocity, and velocity dispersion profiles measured directly from the simulations. By integrating the resulting orbital decay equation and comparing the predicted trajectories with the measured clump orbits, we assess the applicability of the classical dynamical friction formalism and quantify its accuracy and limitations in self-consistent galactic disk environments.

\begin{figure*}
    \centering
    \includegraphics[width=1.7\columnwidth]{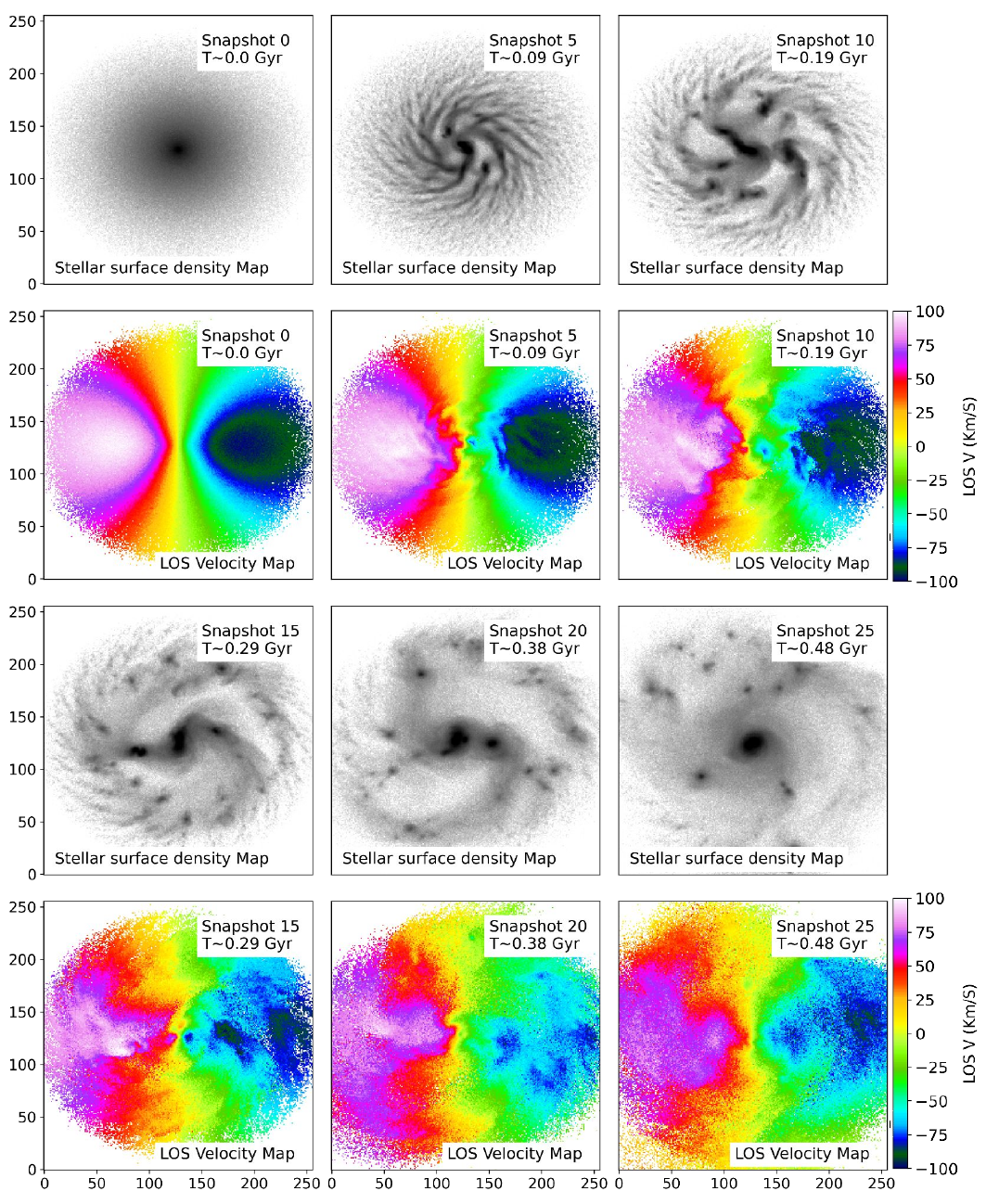}
    \caption{\textbf{Row 1 and Row 3:} Stellar surface density maps for first 25 $(\sim0.5$ Gyr) snapshots showing the transient phase of the galaxy (upto snapshot 20), and a post transient phase disk at snapshot 25, along with disk fragmentation and clump formation. \textbf{Row 2 and Row 4:} The Line of sight velocity maps corresponding to the sellar maps in row 1 and 3. }
    \label{fig:VDI_phase}
\end{figure*}

The paper is organized as follows. In Section~2, we review the theory of dynamical friction. Section~3-4  describes the numerical simulations used in this work and the initial transient phase of the galaxy, while Section~5 presents the modeling of the galaxy rotation curve and dark matter halo. The clump identification procedure is described in Section~6, followed by the clump-tracking algorithm and its limitations in Section~7. In Section~8, we compare the observed evolution of the clumps with the predictions of the classical dynamical friction model. Section~9 discusses the implications of our results for clump migration and the secular evolution of disk galaxies, and Section~10 summarizes our main conclusions.

\section{Dynamical Friction and Inspiral timescale} 
\label{sec:DF}

The classical expression for dynamical friction (DF), first derived by \citep{1943Chandrasekhar}, describes the gravitational drag experienced by a massive object moving through a collisionless background of lighter particles. The force is given by

\begin{equation}
\mathbf{F}_{\rm df} =
-4\pi G^2 M_c^2 \rho \ln\Lambda
\left[
\frac{\mathrm{erf}(X)-\frac{2X}{\sqrt{\pi}}e^{-X^2}}
{V^2}
\right]
\hat{\mathbf{V}},
\label{eqn:Dynamical_fric}
\end{equation}

where $G$ is the gravitational constant, $M_c$ is the mass of the perturber (the clump in our case), $\rho$ is the density of the background medium, and $\ln\Lambda$ is the Coulomb logarithm, which accounts for the cumulative contribution of gravitational encounters over a range of impact parameters. Here, $V$ is the velocity of the clump relative to the background particles, ($\hat{\mathbf{V}}$) is the unit vector in the direction of motion, and
$
X=\frac{V}{\sqrt{2}\sigma},
$
where $\sigma$ is the one-dimensional velocity dispersion of the background. The negative sign indicates that the force acts opposite to the direction of motion. Since the magnitude of the force scales as $M_c^2$, dynamical friction is particularly effective for massive structures such as giant molecular clouds, stellar clumps, satellite galaxies, and galactic bars, causing them to lose orbital energy and angular momentum over time.

For a clump of mass $M_c$ on a circular orbit of radius $R$ with circular velocity ($V(R)$), the orbital angular momentum is

\begin{equation}
L_c=M_cRV.
\end{equation}

The corresponding torque exerted by dynamical friction is

\begin{equation}
\tau=\frac{dL_c}{dt}=RF_{\rm df},
\label{eqn:tau}
\end{equation}

which relates the loss of angular momentum to the radial decay of the orbit. The time required for a clump to migrate from an initial radius $R_a$ to a smaller radius $R_b$ can therefore be written as

\begin{equation}
\Delta T_{\rm insp}^{\rm DF}(R_a\rightarrow R_b)
=
\int_{R_a}^{R_b}
\frac{1}{RF_{\rm df}}
\frac{d(M_cRV)}{dR}
dR
\label{eqn:inspiral_time}
\end{equation}

Assuming that the clump mass remains constant during its evolution, the circular velocity is approximately constant over the interval $[R_b,R_a]$, $V>>\sigma$ in the same interval, and under the assumption of a pseudo isothermal $\rho$, with core radius $R_c$, Equation~\ref{eqn:inspiral_time} simplifies to the following,

$$
\Delta T_{\rm insp}^{\rm DF}(R_a\rightarrow R_b)\propto
\frac{V^3}{M_c}(R_a^2-R_b^2+2R_c^2\ln\frac{R_a}{R_b}).
$$

Thus, the migration time decreases with increasing clump mass and increases with the characteristic rotation velocity of the host galaxy, implying that more massive clumps lose angular momentum more efficiently and spiral toward the galactic center more rapidly. The Chandrasekhar formalism relies on several simplifying assumptions, including that the perturber behaves as a massive point particle moving on a circular orbit through a homogeneous, stationary background with an isotropic Maxwellian velocity distribution. In reality, giant stellar clumps are extended, self-gravitating systems that continuously exchange mass and angular momentum with their surroundings through tidal interactions, stellar feedback, and encounters with other clumps \citep{2007Esquivel,2007Fellhauer,2015Petts}. Consequently, deviations from the classical prediction are expected in realistic galactic environments, which motivates a direct comparison between the analytical formalism and self-consistent numerical simulations.

\begin{figure*}
    \centering
    \includegraphics[width=2\columnwidth]{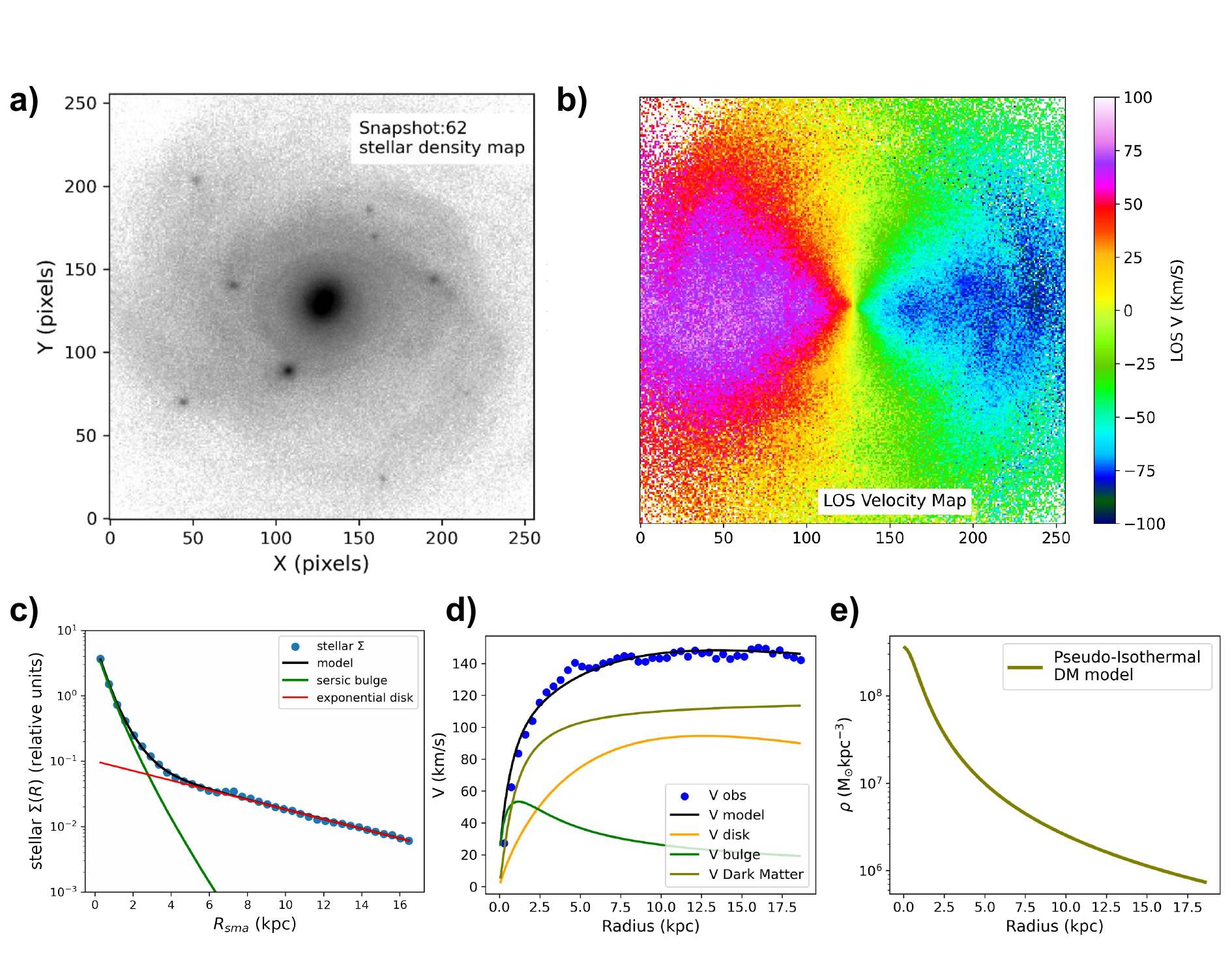}
    \caption{\textbf{(a)} A preview of stellar map of snapshot 62 (marking 1.2 Gyr after the onset of simulation). \textbf{(b)} The Line of sight stellar velocity map corresponding to the stellar map shown in (a). \textbf{(c)} Radial stellar surface density profile of snapshot 62, along with the Sersic bulge and exponential disk model. \textbf{(d)} The rotational velocity curve, along with model disk, bulge, and Isothermal dark matter rotation curves. \textbf{(e)} Density profile of the Dark matter halo inferred from the model dark matter rotation curve in (c).}
    \label{fig:decomp}
\end{figure*}

\section{The Isolated Galaxy Simulations} 
\label{sec:simulations}
The isolated galaxy simulations used in this work are described in \cite{Saha_Cortesi_2018}. Briefly, the initial models comprise three components: a stellar disk, a classical bulge, and a dark matter halo. Each of these components are modelled with a distribution function in steady state, for details on constructing the models, the readers are referred to \cite{KuijkenDubinski1995, Sahaetal2010}. The total stellar mass is $\sim 4.7 \times 10^{10}\ M_\odot$, while the dark matter halo mass is $8.4 \times 10^{10}\ M_\odot$. 
The galaxies are initially disk-dominated and constructed in dynamical equilibrium with a total of $3.7\times10^{6}$ particles. They are evolved for 4.3 Gyr to investigate the secular transformation of disk galaxies into S0-like systems. A suite of four simulations was performed, with initial Toomre stability parameters of $Q = 0.17$, 0.42, 0.73, and 0.94, computed at $2.5 R_d$, $R_d$ being the scale radius. In this study, we focus on the simulation with $Q = 0.73$. The simulation output consists of 220 snapshots; successive snapshots are separated by approximately $19.46$~Myr. For each snapshot, synthetic observations in the form of two-dimensional stellar surface density maps, line of sight (LOS) velocity maps, and LOS velocity dispersion maps were available.
These maps are constructed at an inclination angle of $30^\circ$ and are sampled on a $256 \times 256$ pixel grid with a spatial scale of 0.146 kpc pixel$^{-1}$. 

\section{Initial evolution of the clumpy disk}
\label{sec:initial}
The initial galaxy model is gravitationally unstable to axisymmetric perturbations, with a minimum Toomre parameter of $Q\simeq 0.73$ \citep{1964toomre}. Consequently, the stellar disk rapidly fragments and evolves through a transient clumpy phase during the first $\sim 0.4$ Gyr of the simulation. By $t\sim0.1$ Gyr, the initially smooth stellar disk develops a flocculent, spiral-like morphology (Figure~\ref{fig:VDI_phase}, Snapshot 5). As these spiral features grow non-linearly, they fragment into numerous overdense structures that subsequently evolve into protoclumps. These protoclumps are short-lived, often non-spherical in shape, gravitationally unstable condensations that experience frequent mergers, tidal interactions, and disruption before evolving into the longer-lived clumps analysed in this work.

Clumpy galaxies are commonly observed at high redshift \citep{Kalitaetal2025,delaVegaetal2026}, with recent JWST/NIRCam observations providing new insights into their prevalence and properties. However, the formation, evolution, and survival of giant clumps remain active areas of investigation \citep{2009Dekel,2009Agertz_Teyssier,2017Mandelkar}. In our simulations, clump formation arises solely from classical gravitational instability in a stellar disk with $Q<1$. Although this represents a simplified scenario that does not include additional physical processes such as gas dynamics, star formation, and stellar feedback, it provides a controlled framework for studying the gravitational processes associated with the formation and subsequent evolution of stellar clumps.

\begin{figure*}
    \centering
    \includegraphics[width=2\columnwidth]{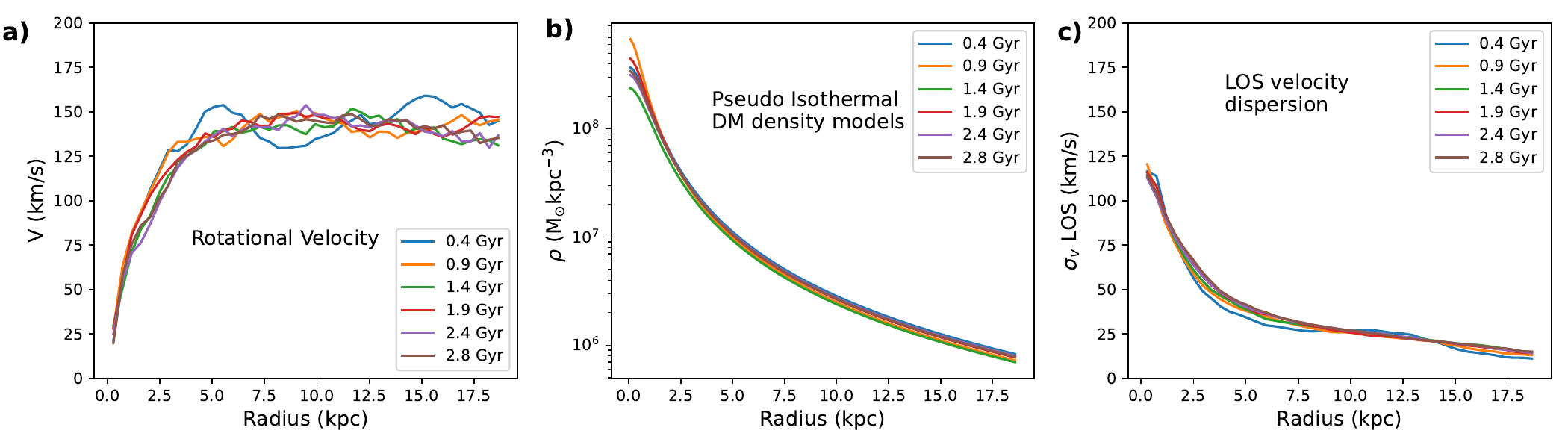}
    \caption{\textbf{(a)} The inclination corrected velocity rotation curve for a number of simulation snapshots from t=0.4 to t=2.8 Gyr. \textbf{(b)} Dark matter density model for the same snapshots as shown in the left panel.
     \textbf{(c)} The Line of sight velocity dispersion of stellar maps as observed from the $\sigma_v$ maps from the simulation.
    }
    \label{fig:all_density}
\end{figure*}

The early evolution of the unstable disk is characterized by rapid structural transformation, accompanied by strong clump--clump interactions and mergers. During this phase, the disk contains a high density of short-lived overdense structures, while the most massive clumps and spiral-arm fragments undergo significant inward radial migration and merge in the central regions (see Snapshots 5 - 25 in Figure~\ref{fig:VDI_phase}). These mergers contribute to the early assembly of the central bulge, increasing the bulge-to-total stellar mass ratio from an initial value of $B/T\sim0.03$ to $B/T\sim0.37$ by the end of the simulation \citep{Saha_Cortesi_2018}. The massive elongated clumps exhibit strong radial motions during this phase, reflecting the highly perturbed nature of the evolving disk.

The rapid structural evolution of the disk is also reflected in its kinematics. The presence of spiral structures and massive clumps produces significant local perturbations and associated streaming motions to the underlying velocity field. Throughout these initial transient phases, one can see various substructures in the velocity maps. For example, at the location of (x,y)=170, 140 in snapshot 15 (Figure~\ref{fig:VDI_phase}), we see an inward migrating clump with a $V_{LOS}\sim 80$~kms$^{-1}$ whereas the local velocity at that location is about $40 - 50$~kms$^{-1}$. In most cases, these migrating clumps seem to produce local streaming motion. Such velocity offsets indicate strong localized deviations from circular rotation produced by the dynamically evolving clump population.

As the disk evolves, the frequency of clump mergers decreases and the system gradually transitions toward a quasi-steady, yet persistently clumpy, configuration by $t\sim0.4$--0.5 Gyr (e.g., Figure~\ref{fig:VDI_phase}, snapshot 25). During this transition, the clump orbits become progressively more circularized \citep{2020Bonetti_circularization} and the velocity field develops a more regular rotational pattern (see Figure~\ref{fig:decomp} at a later stage). Although localized velocity perturbations remain at the positions of individual clumps, their amplitudes decrease as the disk evolves. At $t\sim0.4$ Gyr, these perturbations are typically $\Delta V_{\rm rot}\sim20$--30 km s$^{-1}$, declining to less than $\sim10$ km s$^{-1}$ by $t\sim0.9$ Gyr. The stabilization of the velocity field is accompanied by the convergence of the galaxy rotation curves (see Figure~\ref{fig:all_density}) with reduced radial fluctuation), indicating that the global disk structure has reached an approximate state of dynamical equilibrium while maintaining a clumpy morphology. The orbital decay of clumps during the initial fragmentation phase is influenced by rapid changes in the global potential, frequent clump mergers, and strong non-axisymmetric torques. To minimize these effects and isolate the secular evolution of long-lived clumps, we restrict the analysis to Snapshot 20 and later ($t\gtrsim0.4$ Gyr).

\begin{figure*}
    \hspace{-3cm}
    \includegraphics[width=2.7\columnwidth]{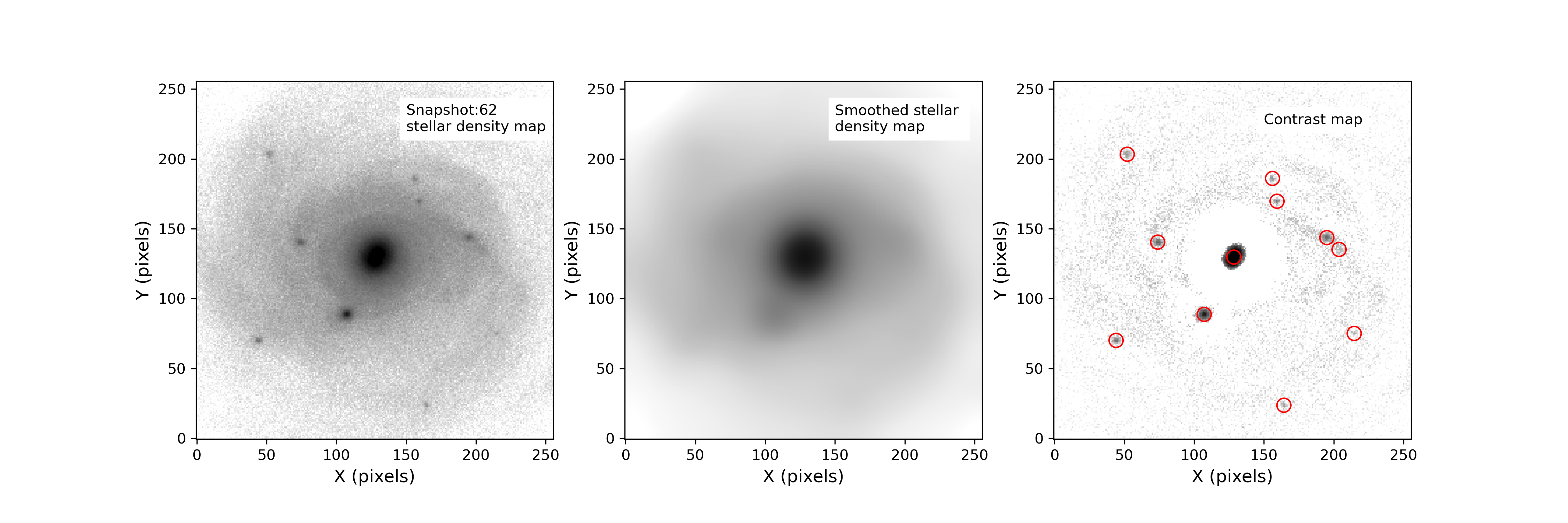}
    \caption{Stellar density map, smoothed stellar density map and Contrast map for the snapshot S=62 ($\sim$ 1.2 Gyr), with detected clumps. The radius of each red circle is 5 pix.}
    \label{fig:contrast}
\end{figure*}

\section{Rotation Curve Decomposition and Isothermal Halo Modeling }
\label{sec:Decomp}
To construct the mass model for each simulation snapshot, we begin with the projected stellar surface density, line-of-sight (LOS) velocity, and LOS velocity dispersion maps.  The LOS velocity is sampled along the semi-major axis. Since the inclination of the simulated galaxy is known ($i=30^\circ$), the observed LOS velocities are corrected for projection effects to obtain the intrinsic circular rotation curve

\begin{equation}
    V_{\rm obs}(R)=\frac{V_{\rm LOS}(R)}{\sin i}
\end{equation}

The projected stellar surface density profile is extracted using elliptical annuli and decomposed into bulge and disk components by fitting a one-dimensional Sérsic profile together with an exponential disk, respectively (example figure \ref{fig:decomp}(b)),

\begin{equation}
    \Sigma(R)=\Sigma_{\rm bulge}(R)+\Sigma_{\rm disk}(R)
\end{equation}

The exponential disk parameters are used to compute the circular velocity contribution of the stellar disk, ($V_{\rm disk}(R)$), assuming an axisymmetric thin exponential disk \citep{1970Freeman,2008Binney_tremaine}. The fitted Sérsic profile is subsequently deprojected under the assumption of spherical symmetry using the analytical approximation of \cite{1997Prugniel_simien}, yielding the three-dimensional bulge density profile. The enclosed bulge mass is then calculated by direct integration.
from which the bulge circular velocity, $V_{\rm bulge}(R)=\sqrt{\frac{GM_{\rm bulge}(<R)}{R}}$ is obtained.

In the simulation set up, the dark matter halo is modelled with a Lowered Evans model \citep{Evans1993} which produces a nearly flat circular velocity profile. Motivated by this and at the same time, treating the simulated velocity maps as it they are from observations, we use a spherical pseudo-isothermal density profile for the dark matter halo:

\begin{equation}
    \rho_{\rm DM}(r)=
\frac{\rho_0}
{1+\left(R/R_c\right)^2},
\end{equation}

\noindent where $\rho_0$ is the central halo density and $R_c$ is the core radius. The corresponding halo rotation curve is

\begin{equation}
   V_{DM}^2= 4\pi G\rho_0r_c^2
\left[
1-
\frac{R_c}{R}
\tan^{-1}\left(\frac{R}{R_c}\right)
\right]
\end{equation}

\noindent The total model rotation curve is constructed by adding the individual contribution from each component in quadrature,

\begin{equation}
    V_{\rm model}^2(R)=V_{\rm bulge}^2(R)
+
V_{\rm disk}^2(R)
+
V_{\rm DM}^2(R).
\end{equation}

\noindent The pseudo-isothermal halo parameters, $\rho_0$ and $r_c$, are determined by fitting $V_{\rm model}(R)$ to the inclination-corrected rotation curve, $V_{\rm obs}(R)$, using a non-linear least-squares minimization (example figure \ref{fig:decomp}c). The resulting best-fit halo parameters are subsequently used to construct the local dark matter density profile required for the dynamical friction calculations (example figure \ref{fig:decomp}d). The rotational velocity curves, best fit Pseudo-Isothermal DM halo profiles, and LOS velocity dispersion for a range of snapshots are shown in figure \ref{fig:all_density}.

\section{Clump Detection} 
\label{sec:clump_detection}
To identify clumps in the simulated stellar surface density maps, we first construct a contrast map by subtracting a Gaussian-smoothed version of each stellar surface density map from the original map. This high-pass filtering technique suppresses the large-scale disk component while enhancing compact overdensities, and is analogous to the methods commonly employed for clump detection in HST and JWST observations \citep{2015Guo,2019Calabro,2024Kalita}. Visual inspection of stellar surface density maps, indicates that the clumps have characteristic radii of approximately 5 pixels. We therefore adopt a Gaussian smoothing kernel with $\sigma=10$ pixels to remove structures on larger spatial scales while preserving the clumps.

Clumps are subsequently identified using SExtractor \citep{1996Sextractor}. Source detection is performed on the contrast maps using an RMS map generated by SExtractor, with a detection threshold of $1.5\sigma$ and a minimum detection area of 10 connected pixels. All other SExtractor parameters are kept at their default values. This procedure is applied to all simulation snapshots spanning the clumpy phase of the galaxy, up to snapshot 180 ($t\approx3.4$ Gyr), after which the disk no longer exhibits a population of giant clumps. An example of a stellar surface density map, smoothed stellar surface density map, and the resulting contrast map along with detected clumps is shown in fig \ref{fig:contrast}. 

\subsection{Clump photometry and mass measurement}
\label{sec:clump_photometry}
\begin{figure}
    \hspace{-1.2cm}
    \includegraphics[width=1.2\columnwidth]{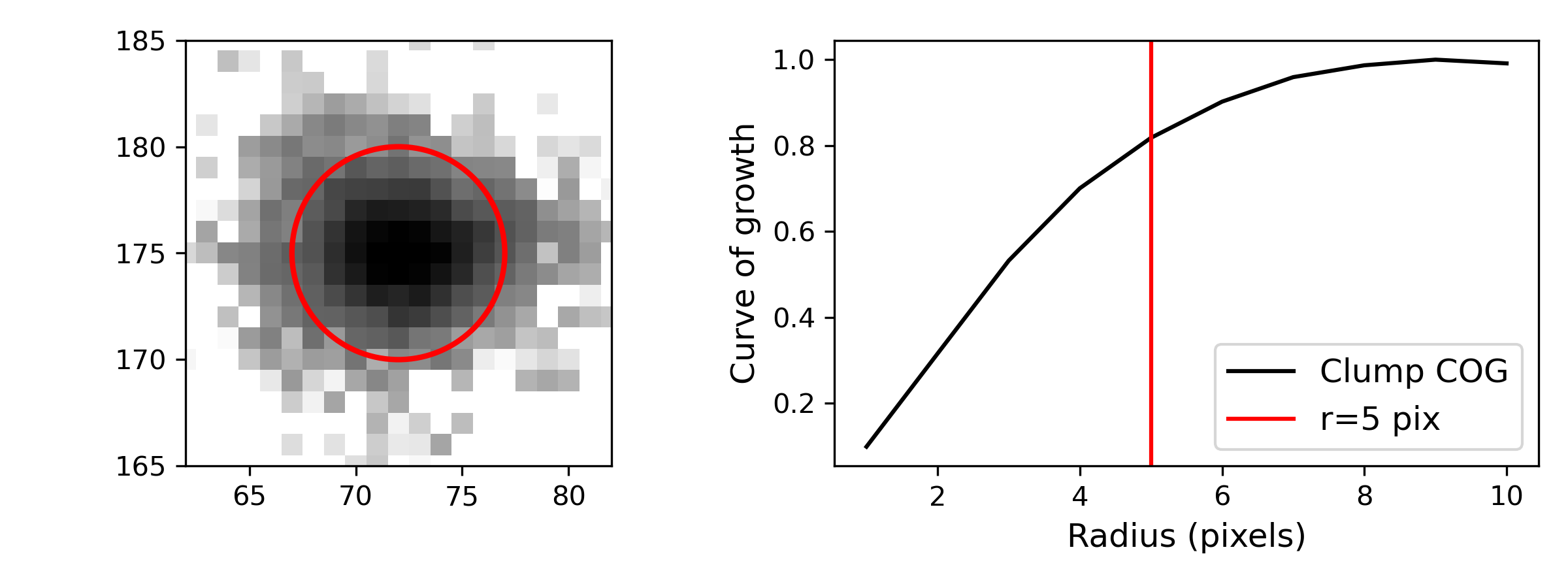}
    \includegraphics[width=1.1\columnwidth]{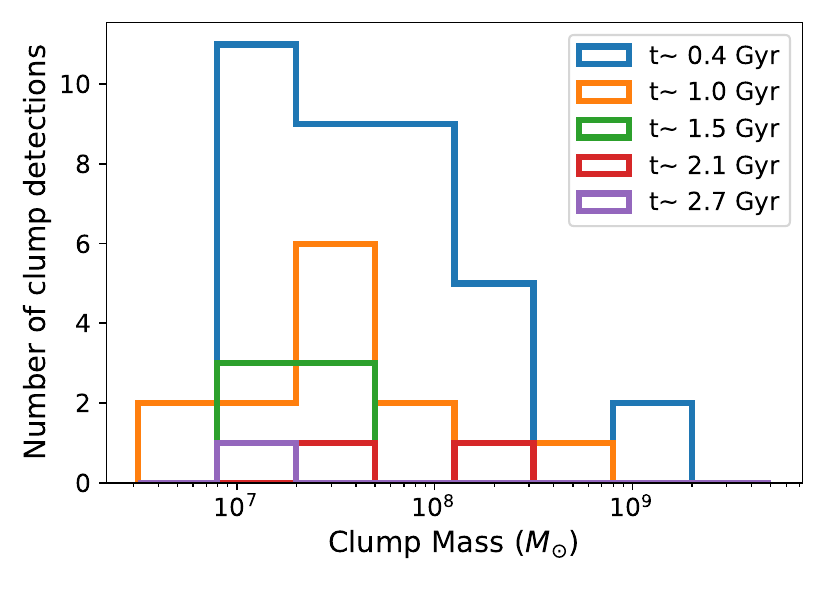}
    \caption{\textbf{Upper panel:} Zoom in view of the most massive long-lived clump (tracking id $20\_36$), and the curve of growth of the clump at snapshot 20. \textbf{Bottom panel:} Mass distribution of detected clumps across simulation snapshots. }
    \label{fig:clump_cog}
\end{figure}

The photometric properties of each clump are measured on the original stellar surface density map using a circular aperture of radius 5 pixels centered on the detected clump positions. 
This standard aperture size was selected by drawing a curve of growth for the most massive long-lived clump (Tracking id `20\_36', as presented in next section), and selecting the radius containing $\gtrsim 80\%$ flux (Figure \ref{fig:clump_cog}).
To account for the local stellar background, we estimate the median background level within a concentric annulus with inner and outer radii of 7 and 10 pixels, respectively, and subtract the corresponding background contribution from the aperture measurement. This local disk background subtraction minimizes contamination from the underlying stellar disk and any nearby diffuse emission. The clump mass distribution for some snapshots are shown in the bottom panel of Figure~\ref{fig:clump_cog}. From the histograms, it is apparent that massive clumps are abundant during the initial phase and their number steadily decreases as the disk evolves.

\begin{figure}
    \centering
    \includegraphics[width=1\columnwidth]{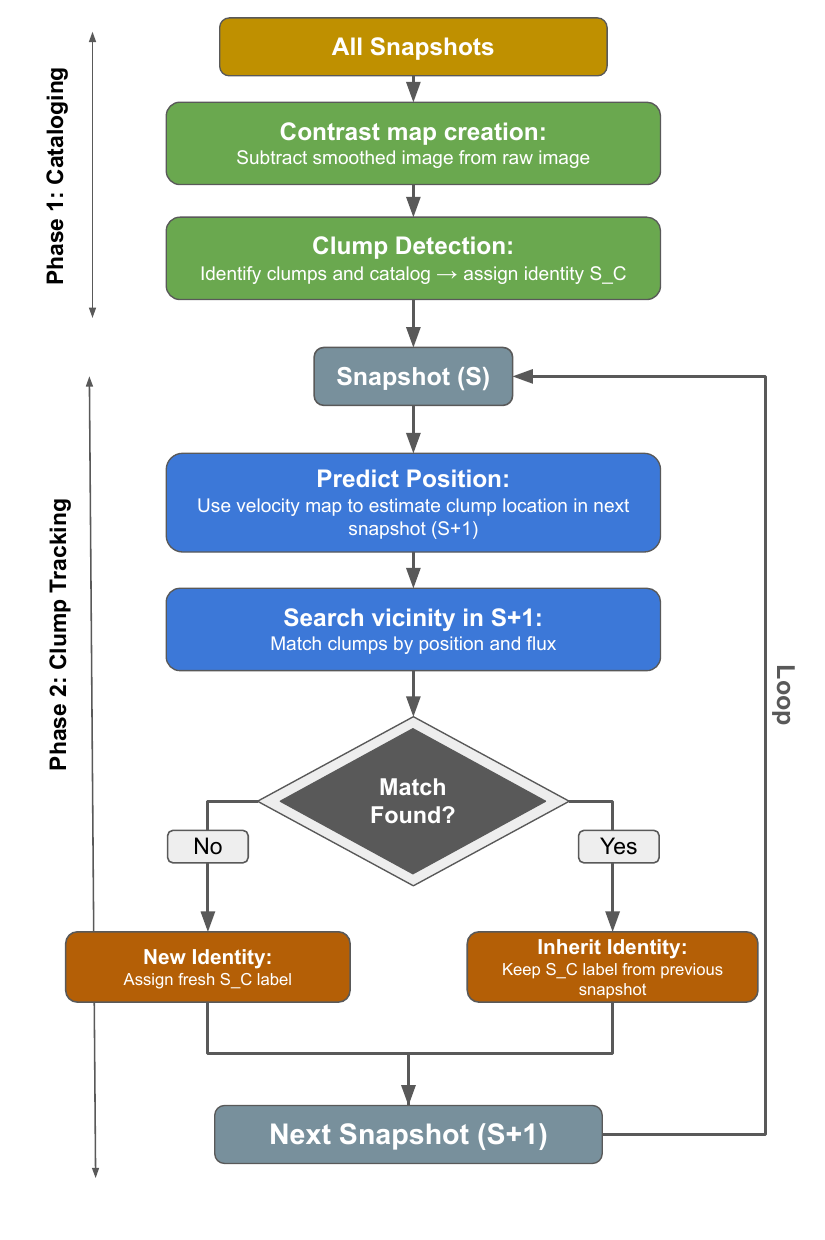}
    \caption{Flow chart summarising the clump tracking process.}
    \label{fig:clump_track_flow_chart}
\end{figure}

\begin{figure*}
    \centering
    \includegraphics[width=2\columnwidth]{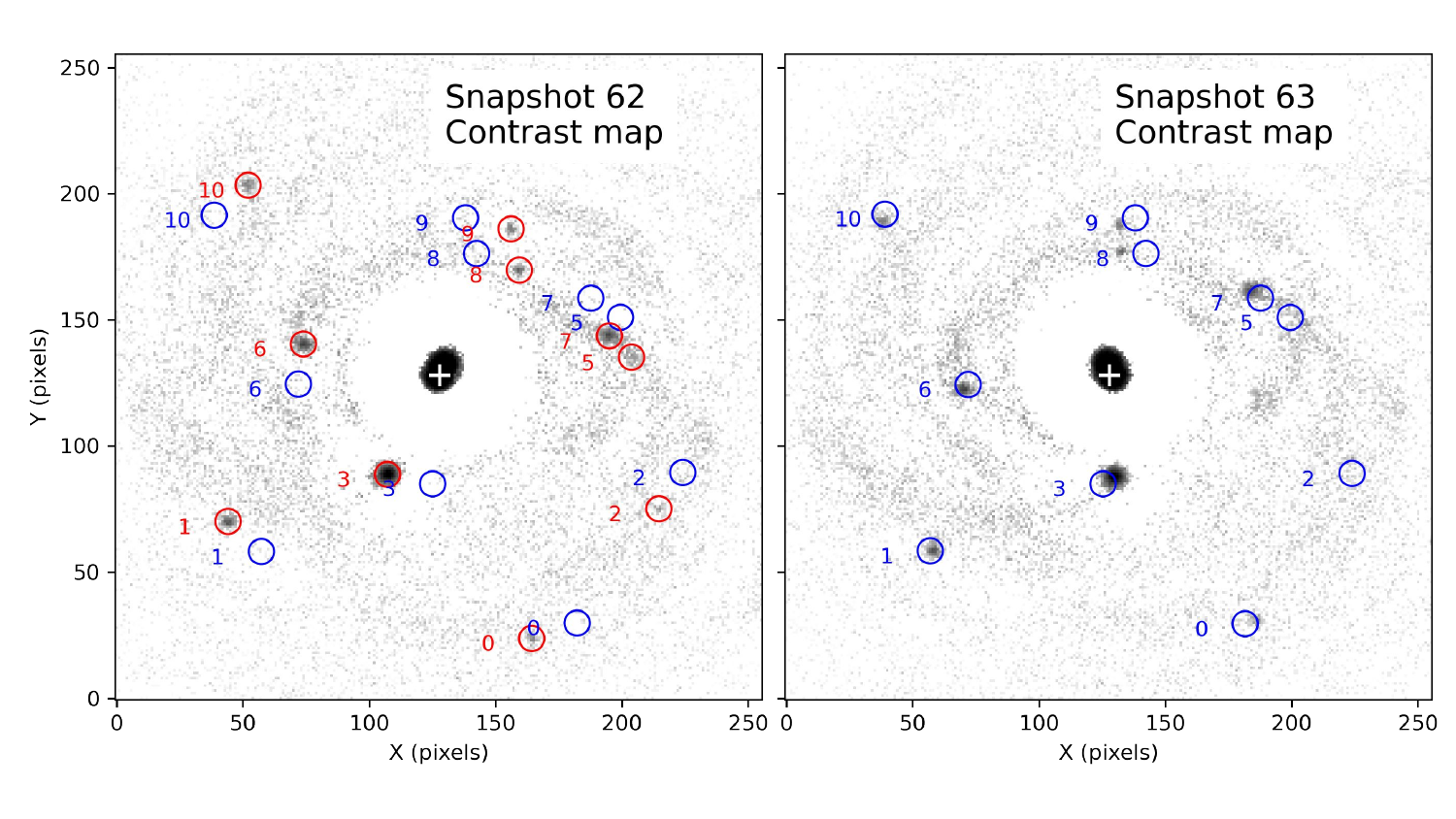}
    \caption{\textbf{(Left)} Contrast map of snapshot 62 ( $\sim$ 1.2 Gyr), with detected clumps (red apertures), along with their predicted position in next snapshot (S=63), using the model velocity curve in figure \ref{fig:decomp}. The numbers represent the detection id from the clump catalog for snapshot 62, and not the clump ids for tracking. \textbf{(Right)} Contrast map of snapshot 63, over-plotted with predicted clump positions (blue apertures) from snapshot 62, which are later crossmatched with detected clumps from snapshot 63.}
    \label{fig:clump_position_prediction}
\end{figure*}

\begin{figure*}
    \centering
    \includegraphics[width=2\columnwidth]{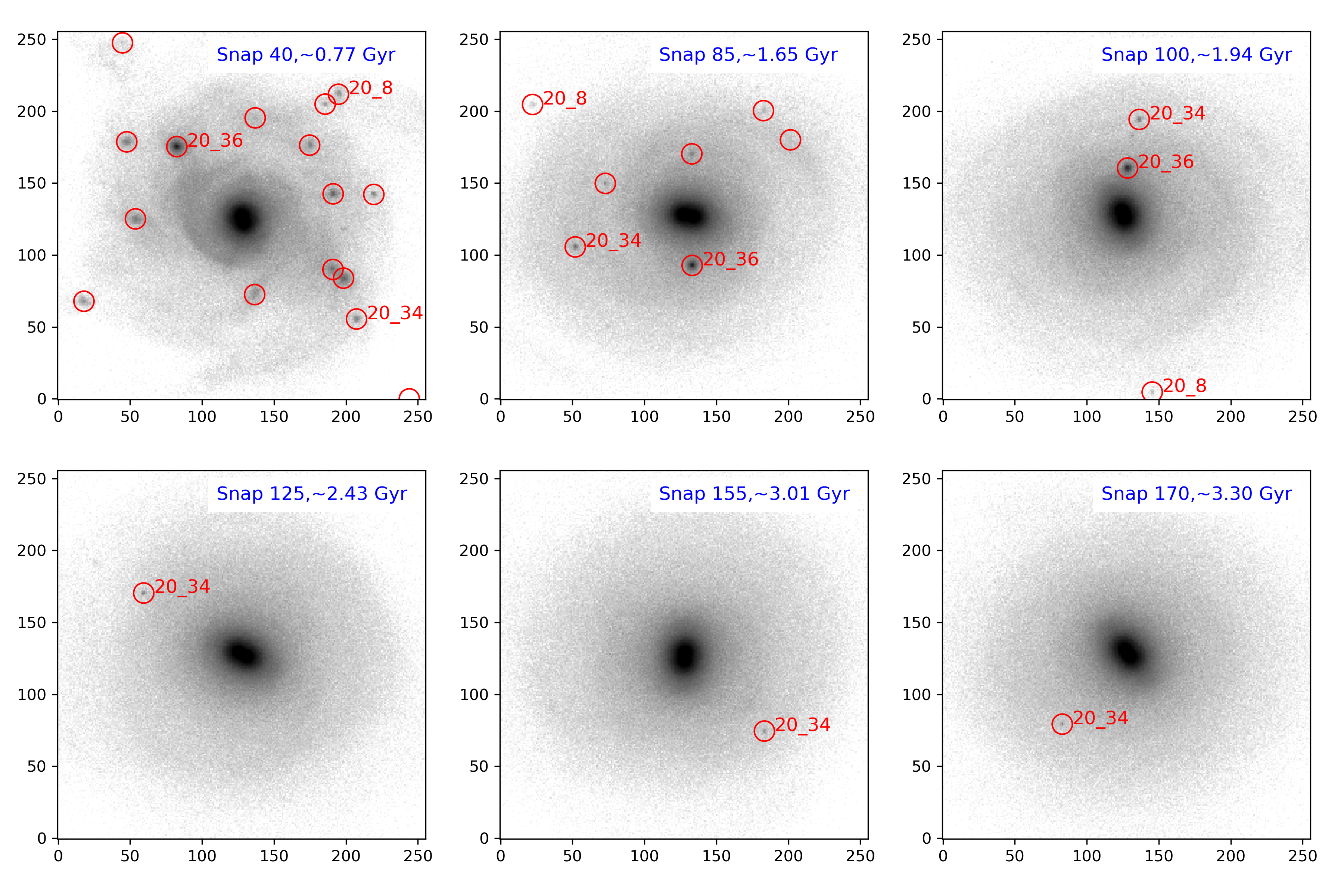}
    \caption{An example of Clump Tracking over a period of $\sim$ 2.5 Gyr. The clump IDs are provided for three long-lived clumps}
    \label{fig:tracked}
\end{figure*}

\section{The Clump Tracking algorithm} 
\label{sec:clump_track}
To measure the inspiral timescale for the individual clumps, they must be tracked throughout the simulation until they either merge with the central bulge or can no longer be traced. For each snapshot, we use the clump catalog, constructed in previous section and match the clumps between consecutive snapshots using the following procedure (also summarised in a flowchart in Figure~\ref{fig:clump_track_flow_chart}):

\begin{enumerate}[label=(\roman*)]
\item A starting snapshot (in our case snapshot 20: t$\sim0.38$ Gyr) is chosen, and all the detected clumps are given an id `S\_C', where S represents the snapshot number, and C represents the detection id in clump catalogs prepared in last section.

\item Using the rotation curve of the galaxy, we estimate the expected position of each clump in the following snapshot, assuming circular motion (example in Figure~\ref{fig:clump_position_prediction}).

\item Clumps are cross matched between consecutive snapshots based on their predicted positions and stellar masses. A maximum mass difference of 0.2 dex, and positional difference of 10 pixels is allowed between the matched clumps. The matched clumps carry on their identity from previous snapshot. Remaining unmatched clumps in the new snapshot are again given a new id S\_C.

\item Repeating this procedure for all snapshots provides the evolutionary track of each clump, including its position and mass as a function of time (Figure~\ref{fig:tracked}).

\end{enumerate}

\subsection{Limitations of the tracking algorithm}
\label{sec:limitations}
Although the tracking algorithm successfully follows most long-lived clumps over multiple snapshots, tracking discontinuities can occur under a few specific circumstances, as summarized below.

\begin{figure*}
    \centering
    \includegraphics[width=2\columnwidth]{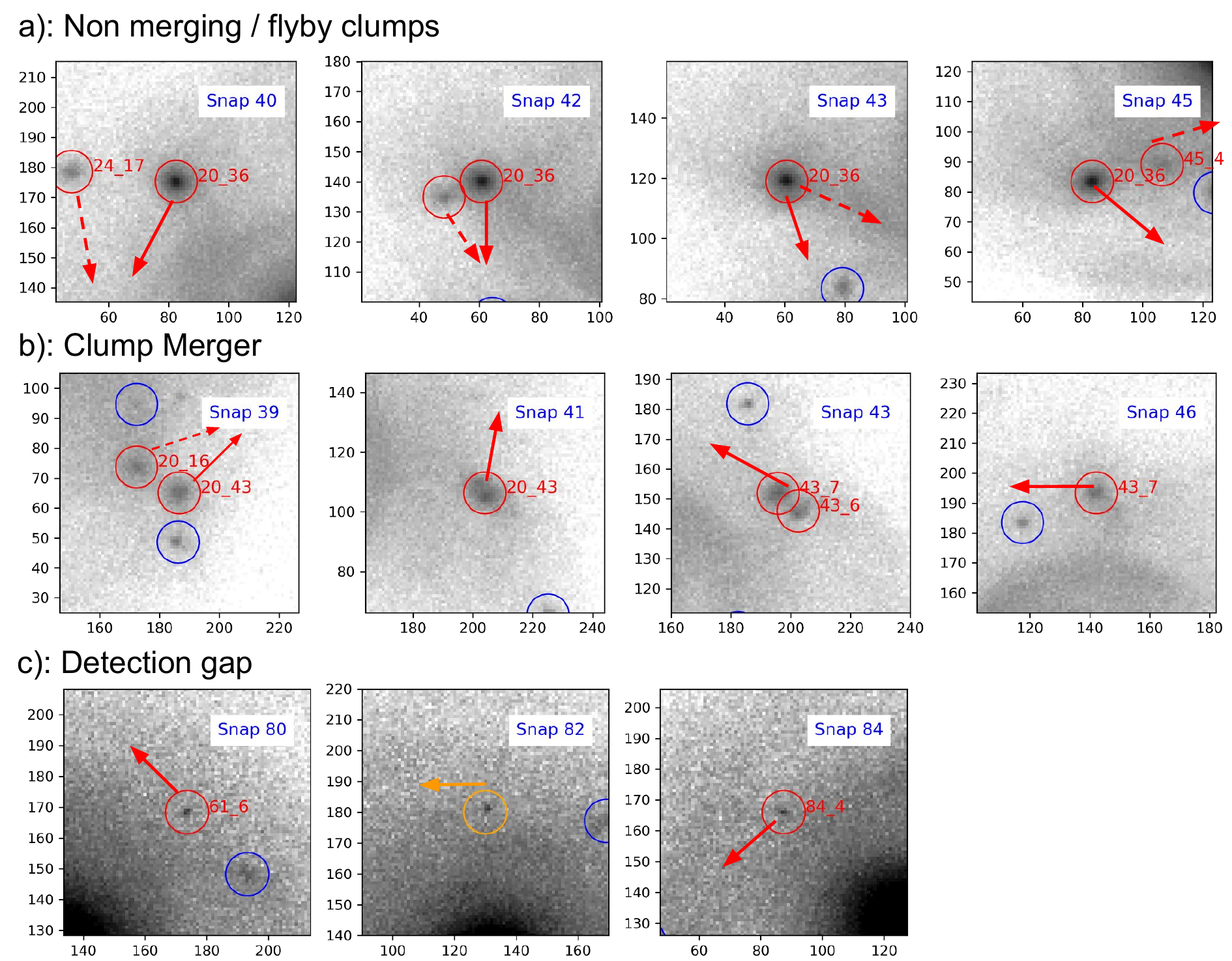}
    \caption{Examples of tracking discontinuities for individual clumps. Clumps of interest are marked with red apertures, while other detected clumps are marked in blue. \textbf{(a)} A non-merging flyby, with solid and dashed red arrows indicating the instantaneous velocities of the two interacting clumps relative to the galactic center. \textbf{(b)} A clump merger, in which the two progenitor clumps approach each other and merge after Snapshot 46; red arrows indicate the instantaneous velocities before and after the merger. \textbf{(c)} A detection gap, in which the clump is not identified in one snapshot (orange aperture), resulting in a discontinuity in the automated tracking.}
    \label{fig:detection_limitation}
\end{figure*}

\begin{figure*}
    \centering
    \includegraphics[width=2\columnwidth]{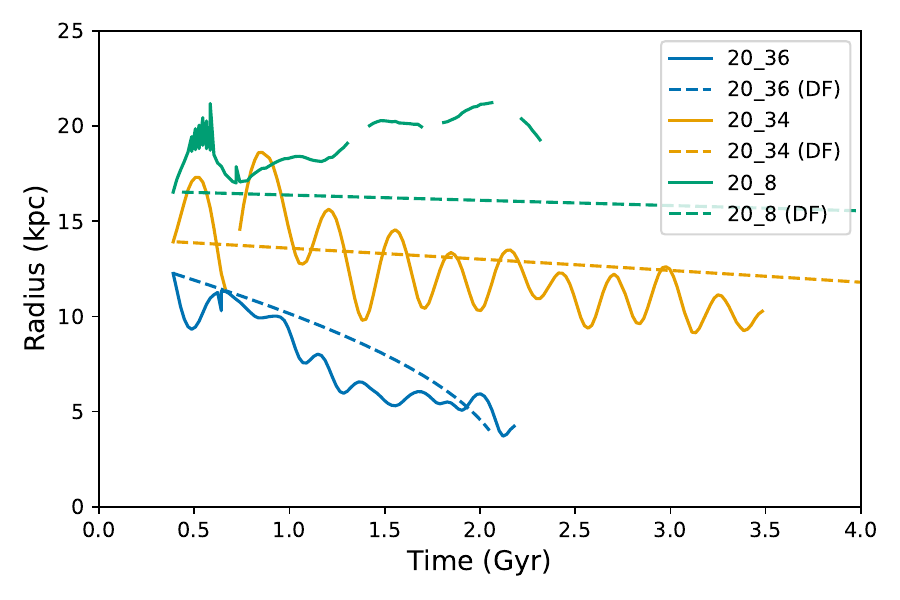}
    \caption{ Time evolution of the galactocentric radius of three representative long-lived clumps directly measured from the simulation (solid lines), together with the corresponding inspiral trajectories predicted by the classical dynamical friction model (dashed lines). The gap in the observed trajectories signify the discontinuities in tracking of the clumps; for example, $20\_8$ moves out of the frame three times, and 20\_34 goes through a non-merging flyby at $t\sim0.6\ Gyr$. }
    \label{fig:T_DFvs}
\end{figure*}

\begin{itemize}
\item \textbf{Non-merging / flyby clumps:} During close encounters, two clumps may overlap in projection without merging. In such cases, the tracking algorithm may temporarily lose one or both clumps and assign a new identity after the encounter.
An example is shown in Figure~\ref{fig:detection_limitation}a, where clumps 20\_36 and 24\_16 undergo a close flyby, during and after which the lower-mass clump is subsequently identified as 42\_7 and 45\_4 respectively.  In rare cases, when both the clumps are identical in mass, the algorithm assigns same id to both.

\item \textbf{Clump mergers:} When two clumps merge into a single structure, the merged object may inherit the identity of one progenitor or be assigned a new identifier. Figure~\ref{fig:detection_limitation}b illustrates such a case, where clumps 20\_16 and 20\_43 merge and are subsequently detected as clump 43\_7. 

\item \textbf{Boundary crossings:} Owing to the finite field of view of the simulation maps, a clump leaving one edge of the image and re-entering from the opposite edge is assigned a new identity upon re-detection. An example is clump 20\_8, which exits the frame and subsequently reappears as 72\_7, before exiting and later re-entering again as 92\_1. These detection gaps are displayed as discontinuities in its path shown in Figure~\ref{fig:T_DFvs}.

\item \textbf{Detection gaps:} Low-contrast clumps may occasionally fall below the detection threshold for one or more consecutive snapshots. Upon re-detection, these clumps are assigned new identifiers. Figure~\ref{fig:detection_limitation}c shows an example where clump 61\_6 is not detected for two consecutive snapshots and later reappears as 84\_4.

\end{itemize}

To minimize the impact of these tracking discontinuities, we visually inspect all clump tracks. Tracks interrupted by non-merging flybys, boundary crossings, or temporary detection gaps are manually linked to recover the complete evolutionary history of the corresponding clumps. In contrast, the merged clumps are treated as newly formed systems and retain their newly assigned identifiers, since their subsequent evolution no longer corresponds to that of either progenitor.

For each tracked clump, the observed inspiral time, $\Delta T_{\rm insp}^{\rm obs}$, is defined as the elapsed time between its first and final detections. The corresponding dynamical friction inspiral time, $\Delta T_{\rm insp}^{\rm DF}$, is obtained by numerically integrating Equation~(\ref{eqn:inspiral_time}) between the same initial and final galactocentric radii. This enables a direct comparison between the inspiral times predicted by the classical dynamical friction formalism and those measured from the snapshots.

\begin{figure}
    \centering
    \includegraphics[width=1\columnwidth]{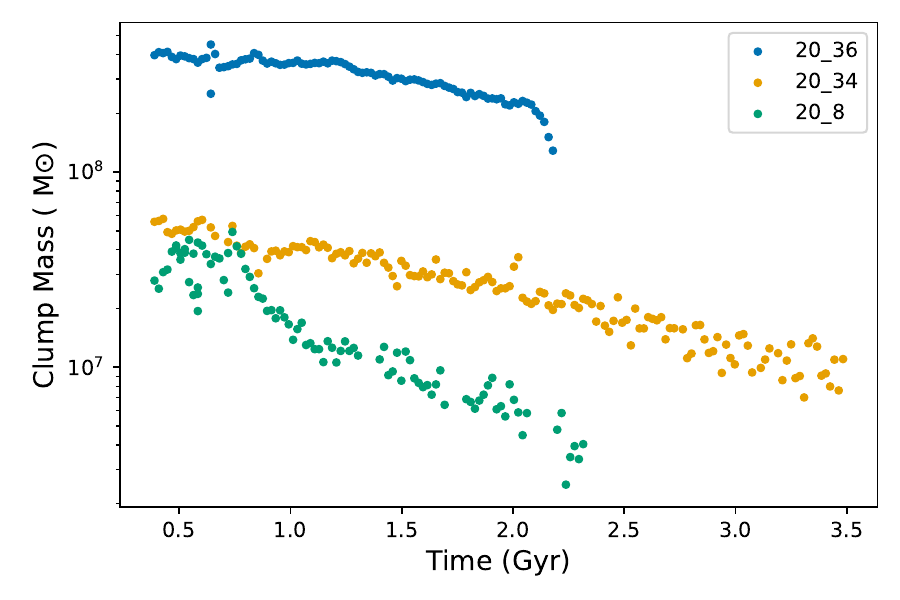}
    \caption{ Time evolution of the masses of three representative clumps during the tracking period. }
    \label{fig:T_DMvs}
\end{figure}

\begin{figure*}
    \centering
    \includegraphics[width=2.\columnwidth]{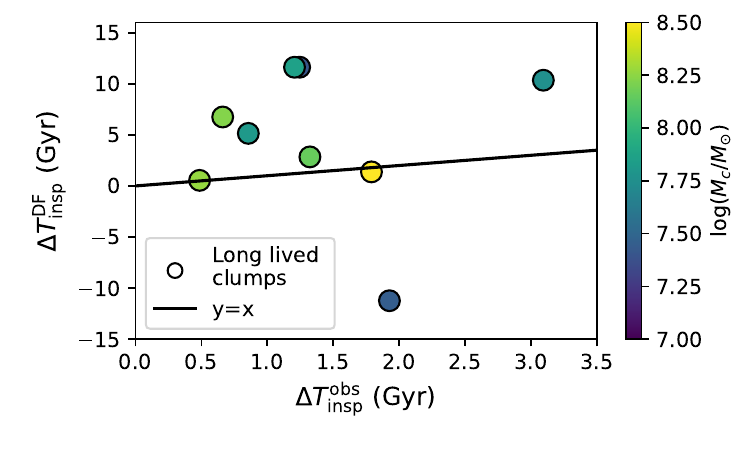}
    \caption{ Comparison between the observed inspiral times, $\Delta T_{\rm insp}^{\rm obs}$, and the corresponding dynamical friction predictions, $\Delta T_{\rm insp}^{\rm DF}$, for all tracked long-lived clumps.}
    \label{fig:DeltaT_comp}
\end{figure*}

\section{Evolution of the clumps and their migration timescales}
\label{sec:Evolution_of_clumps}
During the first $\sim400$ Myr of the simulation, the stellar disk undergoes rapid fragmentation, producing numerous short-lived overdensities (protoclumps) that experience frequent interactions and mergers before evolving into a quasi-steady clumpy disk (Figure~\ref{fig:VDI_phase}). As discussed in Section~\ref{sec:Decomp}, we therefore restrict our analysis to snapshots beyond $T\approx400$ Myr, when the global properties of the galaxy have stabilized.

The tracked clumps follow mildly elliptical orbits about the galactic center, in contrast to the circular-orbit assumption adopted in the classical dynamical friction formalism. Since the orbital period at a typical galactocentric radius of $R\sim10$ kpc is $\approx420$ Myr, robust measurements of orbital evolution require clumps to be tracked for at least one complete orbit. Beyond $\Delta T^{obs}\approx420$~Myr, we identify $9$ long-lived clumps that remain detectable for more than 20 consecutive snapshots (corresponding to $\Delta T^{obs}\gtrsim 400$~Myr). The remaining clumps either dissipate on shorter timescales or undergo complex interactions that prevent continuous tracking; throughout this work, we refer to these as `short-lived clumps'. Nevertheless, most of the long-lived clumps experience substantial mass loss throughout their lifetimes. Figure~\ref{fig:T_DMvs} displays a representative example, illustrating the nearly steady decline in clump mass during their evolution. By the end of their tracked lifetimes, the long-lived clumps have, typically, lost $60$-$90\%$ of their initial mass. 
As per the orbital evolution is concerned, eight of the nine long-lived clumps exhibit a net decrease in galactocentric radius over their observed lifetimes, indicative of orbital decay. Figure~\ref{fig:T_DFvs} compares the measured orbital evolution of three representative clumps with the trajectories predicted by the classical dynamical friction model. The outer-disk clump 20\_8 shows no evidence of inspiral. Instead, it exhibits a gradual outward migration, consistent with the negligible orbital evolution predicted by the dynamical friction model over the duration of the simulation. In contrast, clumps 20\_36 and 20\_34 both migrate inward. The most massive clump, 20\_36, follows the inspiral trajectory predicted by dynamical friction remarkably well before merging with the central bulge, whereas clump 20\_34 deviates from the predicted evolution shortly before dispersing into the stellar disk.

To assess the consistency between the simulations and the analytical model, we compare the inspiral time measured directly from the simulations, $\Delta T_{\rm insp}^{\rm obs}$, with the dynamical-friction inspiral time, $\Delta T_{\rm insp}^{\rm DF}$, computed from Equation~\ref{eqn:inspiral_time} over the same radial interval (Figure~\ref{fig:T_DFvs}). For clump 20\_8, the orbital radius increases over the measurement interval, yielding a negative value of $\Delta T_{\rm insp}^{\rm DF}$ because the integration is performed between an initial and a final radius where latter is larger than the initial one. This behaviour indicates outward migration rather than inspiral and therefore falls outside the regime in which the analytical inspiral timescale is physically meaningful. Among the remaining long-lived clumps, two exhibit inspiral times broadly consistent with the predictions of the dynamical friction model, while six migrate significantly more rapidly than predicted, with $\Delta T_{\rm insp}^{\rm DF}$ exceeding $\Delta T_{\rm insp}^{\rm obs}$ by factors of approximately $2$--$10$, the discrepancy being stronger for the low-mass clumps (see Figure~\ref{fig:DeltaT_comp}). The outward migrating clump 20\_8 appears with a negative $\Delta T_{\rm insp}^{\rm DF}$.  Overall, the Chandrasekhar's dynamical friction model systematically overestimates the inspiral timescale for the majority of long-lived clumps.

An additional point is that the analytical estimates neglect the secular mass loss experienced by the simulated clumps (as discussed above and shown in Figure~\ref{fig:T_DMvs}). Since the Chandrasekhar dynamical friction timescale scales approximately as $\Delta T_{\rm insp}^{\rm DF} \propto M_{c}^{-1}$, the gradual decrease in clump mass would weaken the dynamical friction force and therefore increase the predicted inspiral time. Consequently, accounting for the observed mass evolution would further enlarge the discrepancy between the analytical predictions and the measured migration times. The faster migration observed in the simulations therefore cannot be attributed to neglecting clump mass loss; instead, it indicates that additional mechanisms, beyond classical Chandrasekhar dynamical friction, contribute significantly to the orbital decay.

\section{Discussions}
\label{sec:discussion}
The Chandrasekhar dynamical friction formalism provides a simple analytical framework for estimating the orbital decay of massive objects embedded in a collision-less background. While the predicted inspiral times generally reproduce the overall trend of clump migration, noticeable deviations are present for several individual clumps. These differences are expected, as the assumptions underlying the analytical model are only approximately satisfied in a realistic galactic disk.

The Chandrasekhar formalism assumes that the perturber is a massive point particle moving through an infinite, homogeneous, and stationary background with an isotropic Maxwellian velocity distribution. In contrast, the stellar disk in our simulations is highly non-axisymmetric, exhibiting strong radial density gradients, ordered rotation, finite thickness. As a result, the local environment experienced by a clump evolves continuously during its lifetime, modifying the strength of the dynamical friction force.

\subsection{Sources of uncertainty}
One of the primary sources of uncertainty in the analytical model is the evolution of the clump mass. The classical dynamical friction formalism assumes a constant clump mass, whereas the simulated clumps continuously exchange mass with their surroundings through tidal stripping, mergers, and interactions with the local disk environment. Since the Chandrasekhar drag scales as $F_{\rm DF}\propto M_c^2$, even modest variations in clump mass can significantly affect the predicted inspiral rate. Mass loss reduces the dynamical friction force and consequently lengthens the inspiral timescale, while mass growth has the opposite effect. However, despite experiencing substantial mass loss throughout their evolution, the long-lived clumps in our simulations generally inspiral more rapidly than predicted by the classical dynamical friction model (Figure~\ref{fig:T_DFvs}), which is also seen in literature \citep{2008Fujii_Iwasawa}. This suggests that clump mass evolution alone cannot account for the observed discrepancy between the measured and predicted inspiral times. 

Close encounters between neighbouring clumps provide an additional source of deviations from the smooth dynamical friction picture. During such interactions, clumps can exchange energy and angular momentum, undergo gravitational scattering, or temporarily overlap in projection \citep{2009Dekel,2013Fiacconi_mayer}. These processes introduce changes in clump orbits that are not captured by an analytical dynamical friction prescription. Furthermore, during the initial phase of disk fragmentation, the system undergoes rapid nonlinear evolution, with transient spiral structures and strongly time-dependent gravitational torques. Although we have excluded this early violent phase from our analysis, the disk retains residual non-axisymmetric features at later times, particularly spiral structures (see Figure~\ref{fig:VDI_phase}, snapshot 25). These structures can exert gravitational torques on the clumps and contribute to angular momentum redistribution \citep{LBK1972,SahaJog2014,BanikvandenBosch2021}. In a live disk–halo system, the angular momentum evolution of a clump is therefore determined by the combined effect of several gravitational torques. The total torque can be expressed schematically as 
$$\tau_{\rm tot}=\tau_{\rm disk}+\tau_{\rm halo}+\tau_{\rm int}$$,

\noindent where $\tau_{\rm disk}$ represents the torque from the stellar disk, including contributions from the clump-induced wake and residual non-axisymmetric structures; $\tau_{\rm halo}$ represents the response of the dark matter halo associated with dynamical friction, and $\tau_{\rm int}$ accounts for direct clump–clump interactions. While the Chandrasekhar formalism captures the leading contribution expected from gravitational wakes, the additional torques arising from the live galactic environment introduce deviations from the idealized analytical prediction.

Despite these limitations, the Chandrasekhar formalism reproduces the overall trend of clump migration and its dependence on clump mass and galactocentric radius. This suggests that dynamical friction captures the leading-order behaviour of the secular inspiral of massive stellar clumps. The remaining deviations likely arise from the complex and time-dependent nature of realistic galactic disks, including clump–clump interactions, residual non-axisymmetric structures, and the evolving masses of the clumps themselves.
 
\subsection{Implications for Clump Migration in real disk galaxies}

The isolated galaxy model adopted in this study provides a controlled framework for testing the applicability of classical dynamical friction to the evolution of massive clumps, but it represents a simplified description of real star-forming galaxies. The simulation includes only collisionless stellar and dark matter components and does not account for gas dynamics, star formation, stellar feedback, radiation pressure, or interactions with the multiphase interstellar medium, all of which can influence clump formation, mass evolution, and orbital migration \citep{2010caverino,2010Murray_Quartaert,2012Hopkins_Kere,2017Mandelkar}. Nevertheless, the clumps in our simulations evolve within a self-consistent stellar disk and dark matter halo, undergo interactions with neighbouring clumps, experience substantial mass loss, and follow non-circular orbits, providing a more realistic environment than the idealized static, homogeneous background assumed in the classical Chandrasekhar formalism. Our results therefore demonstrate that classical dynamical friction alone may not provide a reliable estimate of clump inspiral timescales, even in the absence of additional baryonic processes. Consequently, estimates of the amount of mass transported from the outer disk toward the galactic center, and hence the contribution of clump migration to bulge growth, should be treated with caution when based solely on classical dynamical friction. 

\section{Summary and Conclusions}
We have investigated the migration of stellar clumps in an isolated collisionless disk galaxy simulation and assessed the applicability of the classical Chandrasekhar dynamical friction formalism in predicting their orbital evolution. Using a sequence of projected stellar density and kinematic maps, we identified and tracked individual clumps over their lifetimes, constructed baryonic and dark matter mass models for each simulation snapshot, and computed the local dynamical friction force from the measured density, circular velocity, and velocity dispersion profiles. The predicted inspiral times were then compared with the directly measured evolution of the clumps in the simulation.

Our main findings are summarized below.

\begin{enumerate}
\item Following an initial transient phase lasting approximately $\sim400$ Myr, the galaxy settles into a quasi-steady state, characterized by stable rotation curves and nearly unchanged best-fit pseudo-isothermal halo parameters. The subsequent analysis is therefore restricted to this evolved phase.

\item The majority of massive clumps exhibit systematic inward migration toward the galactic centre, whereas lower-mass clumps generally experience longer migration timescales and more irregular orbital evolution.

\item Almost all detected clumps undergo continuous mass loss throughout their migration. The smooth nature of this mass evolution suggests that gradual tidal stripping, rather than stochastic disruption from close encounters, is the dominant mechanism regulating clump mass loss.

\item For the massive clumps tracked over extended periods, the measured inspiral timescales are shorter by a factor of $2$--$10$ compared to predictions based on Chandrasekhar's dynamical friction formalism. These differences indicate that additional physical processes contribute to the orbital evolution beyond the idealized assumptions of the analytical model. Possible contributions include the evolving clump mass, gravitational interactions between neighbouring clumps, departures from circular orbits, and torques generated by the self-gravitating stellar disk. These effects become particularly important for lower-mass clumps and in regions where dynamical interactions are stronger.

\end{enumerate}

\software{
          Source Extractor \citep{1996Sextractor}, Photutils \citep{bradley_2026_19636730}
          }

\bibliography{sample701}{}
\bibliographystyle{aasjournalv7}

\end{document}